\documentstyle[12pt,epsfig]{article}                               
                               
\newlength{\dinwidth}                               
\newlength{\dinmargin}                               
\def\lapproxeq{\lower .7ex\hbox{$\;\stackrel{\textstyle                               
<}{\sim}\;$}}                               
\def\gapproxeq{\lower .7ex\hbox{$\;\stackrel{\textstyle                               
>}{\sim}\;$}}                               
\def\be{\begin{equation}}                               
\def\ee{\end{equation}}                               
\def\bea{\begin{eqnarray}}                               
\def\eea{\end{eqnarray}}    
 
\def\xtil{\tilde{x}}

\begin{document}                               
\titlepage                               
\begin{flushright}                               
DTP/99/24 \\                               
March 1999 \\                               
\end{flushright}                               
                               
\vspace*{2cm}                               
                               
\begin{center}                               
{\Large \bf Diagonal input for the evolution of off-diagonal partons}                               
                               
\vspace*{1cm}                               
K.J.~Golec-Biernat$^{a,b}$, A.D.~Martin$^a$ and M.G.~Ryskin$^{a,c}$ \\                               
                              
\vspace*{0.5cm}                               
$^a$ Department of Physics, University of Durham, Durham, DH1 3LE \\                              
$^b$ H.~Niewodniczanski Institute of Nuclear Physics, ul.~Radzikowskiego 152,                   
Krakow, Poland \\                  
$^c$ Petersburg Nuclear Physics Institute, Gatchina, St.~Petersburg, 188350, Russia                              
\end{center}                               
                               
\vspace*{2cm}                               
                               
\begin{abstract}                               
We show that a knowledge of diagonal partons at a low scale is sufficient to determine 
the off-diagonal (or skewed) distributions at a higher scale, to a good degree of 
accuracy.  We quantify this observation by presenting results for the evolution of 
off-diagonal distributions from a variety of different inputs.
\end{abstract}                              
                      
\newpage                              
\noindent {\large \bf 1.~~Introduction}                
                
Precision data are becoming available for hard scattering processes whose description                      
requires knowledge of off-diagonal (or so-called \lq\lq skewed") 
parton distributions \cite{ROB,RAD1,JI1}.                       
For example the diffractive production of vector mesons or  high $E_T$ jets in                     
high energy electron-proton collisions are now experimentally accessible.                     
                     
Off-diagonal distributions\footnote{Here we use the off-forward distributions with              
support $-1 \leq x \leq 1$ introduced by Ji \cite{JI1}, except that we take $H_g = 
xH_g^{\rm Ji}$.  In the limit $\xi \rightarrow 0$ the distributions reduce to the 
conventional diagonal distributions:  $H_q (x, 0) = q (x)$ for $x > 0$, $H_	q (x, 0) = - 
\bar{q} (-x)$ for $x < 0$ and $H_g (x, 0) = xg (x)$.  See \cite{JI3,RAD2,GM} for 
detailed discussion of off-diagonal distributions.}              
$H (x, \xi, t, \mu^2)$ depend on the momentum fractions $x_{1,2} = x \pm \xi$              
carried by the emitted and absorbed partons at each scale $\mu^2$, see the lower part              
of Fig.~1.  They also depend on the momentum transfer variable $t = (p -              
p^\prime)^2$.  However $t$ does not change as                      
we evolve the parton distributions up in the scale $\mu^2$.  That is $t$ lies outside                      
the evolution and we can study the scale and $x_i$ dependence of $H$ for each fixed                      
value of $t$.  The evolution of the distributions $H$ with the scale is described by                
\be                
\label{eq:a1}                
\mu^2 \frac{\partial H (x, \xi)}{\partial \mu^2} \; = \;\frac{\alpha_S}{2 \pi} \int  
dx_1^\prime dx_2^\prime \: \delta (x_1 - x_2 - x_1^\prime + x_2^\prime) \: P  
(x_1^\prime, x_2^\prime, x_1, x_2) \: H (x^\prime, \xi^\prime),                
\ee                
see Fig.~1, where the off-diagonal splitting functions $P$ are known to leading order 
(LO) and the anomalous dimensions are known to next-to-leading order (NLO) 
accuracy.  So there is a possibility to evolve $H$ numerically starting from some input 
distribution. 
                
As in the diagonal case, we require the form of the starting distributions, which will                
be dependent on non-perturbative QCD.  This poses a major problem for off-diagonal                 
evolution since we must specify both the $x_1$ and $x_2$ behaviour of the starting                 
distributions.  We could, in principle, determine the off-diagonal input directly from                 
data.  However the possibility of determining the $x_1$ and $x_2$ behaviour, for a                 
given $t$ value, from data is remote.  Alternatively we could use some simplified                 
model for the distributions in the non-perturbative region.  Clearly it would be better                
to try to relate the off-diagonal distributions to their known diagonal form. 
 
In the present paper we argue that at a low scale $\mu^2 = Q_0^2$ one may input the  
diagonal forms, but smoothed in the ERBL region with $| x | < \xi$ to avoid the  
singularity at $x \rightarrow 0$.  Then after a few steps of evolution the result will be  
very close to the true off-diagonal distribution.  In other words the original  
$\xi$ dependence is washed out and forgotten during the evolution, with the final  
$\xi$ behaviour being generated mainly by the form of the off-diagonal splitting  
functions. 
 
To some extent, the idea to use a pure diagonal input is supported by the GRV  
approach \cite{GRV}.  With more or less simple, valence-like distributions at very  
low input scale $Q_0^2$, the evolution generates quite reasonable diagonal parton  
distributions, whose $x$-dependence is essentially governed by the form of the  
evolution equation. 
 
In the next section we give a better argument in favour of diagonal input at a low scale  
for generating reliable off-diagonal distributions by evolution.  The argument is based  
on the polynomial properties of the operator product expansion (OPE) for the function  
$H (x, \xi)$. 
 
In Section 3 we demonstrate numerically that starting at $Q_0^2 = 0.26$~GeV$^2$  
with different inputs, which have the same diagonal limit for $\xi \rightarrow 0$, we  
already obtain practically the same distribution $H (x, \xi)$ in the region $Q^2 \sim  
2-10$~GeV$^2$.  Thus one can reasonably specify the off-diagonal distributions  
directly in terms of the (conventional) diagonal partons which are known from  
experiment. \\ 
 
\noindent {\large \bf 2.~~Constraints based on the polynomial condition} 
 
First we recall the known properties\footnote{We consider the behaviour at $t = 0$.  
Otherwise we may assume the factorization property $H (x, \xi; t, \mu^2) = G (t) H (x, 
\xi; \mu^2)$, where $G (t)$ is the proton form factor with $G (0) = 1$.} of the 
function $H (x, \xi)$.  Just from the Lorentz invariance and the tensor structure of the 
operators in the OPE, it follows that the  
$x$-moments 
\be 
\label{eq:a2} 
H_n (\xi) \; = \; \int_{-1}^1 \: x^n \: H (x, \xi) \: dx \; = \; \sum_k^{\left [ \frac{n +  
1}{2} \right ]} \: C_{kn} \xi^{2k} 
\ee 
are even polynomials in powers of $\xi$ of order $N \leq n + 1$ (i.e.~$2k \leq n +  
1$) \cite{JI3}.  The odd powers of $\xi$ are absent due to the left-right $(\xi 
\rightarrow  
- \xi)$ symmetry of Fig.~1.  Expanding now the off-diagonal distribution in the  
powers of $\xi^2$ 
\be 
\label{eq:a3} 
H (x, \xi) \; = \; \sum_{k = 0}^\infty \: \xi^{2k} \: h_{2k} (x), 
\ee 
we see that the lower moments of the coefficient functions $h_{2k} (x)$ with $n + 1 
< 2k$  
are required to be zero 
\be 
\label{eq:a4} 
\int_{-1}^1 \: x^n \: h_{2k} (x) dx \; = \; 0. 
\ee 
Therefore the larger the value of $k$ the more the functions $h_{2k} (x)$ must  
change sign and oscillate in the interval [$-1, 1$] in order to satisfy the moment  
conditions.  These oscillations will be washed out and cancel each other during the  
evolution.  Thus the original (input) $\xi$ behaviour will die out as $Q^2$ increases  
and only the $\xi$ dependence generated by evolution will survive. 
 
Another way to arrive at the same conclusion is to consider the properties of the  
anomalous dimensions $\gamma_n$.  Here it is better to consider the conformal  
moments $O_n$ \cite{OHRNDORF} 
\be 
\label{eq:a5} 
O_n^i \; = \; \int \: R_n^i (x_1, x_2) \: H_i (x, \xi) \: dx, 
\ee 
with $i = q, g$, defined on the polynomial bases 
\bea 
\label{eq:a6} 
R_n^q (x_1, x_2) & = & \sum_{k = 0}^n \left ( \begin{array}{c} n \\ k \end{array} 
\right ) \: \left ( \begin{array}{c} n + 2 \\ k + 1 \end{array} \right ) \: x_1^k \: x_2^{n - 
k} \\ 
& & \nonumber \\ 
\label{eq:a7} 
R_n^g (x_1, x_2) & = & \sum_{k = 0}^n \left ( \begin{array}{c} n \\ k \end{array} 
\right ) \: \left ( \begin{array}{c} n + 4 \\ k + 2 \end{array} \right ) \: x_1^k \: x_2^{n - 
k}  
\eea 
with $x_{1,2} = x \pm \xi$.  The conformal moments have the advantage that they do 
not mix during LO evolution, and are simply renormalized multiplicatively 
\be 
\label{eq:a8} 
O_n (Q^2) \; = \; O_n (Q_0^2) \: \left ( \frac{Q^2}{Q_0^2} \right )^{\gamma_n}. 
\ee 
They satisfy the same polynomial properties as the $x$ moments of (\ref{eq:a2}).  So 
only the anomalous dimensions  
$\gamma_n$ with $n + 1 > 2k$ may participate in the evolution of the coefficient  
function $h_{2k} (x)$.  On the other hand the values of $\gamma_n$ decrease with  
increasing $n$.  In fact $\gamma_n$ become negative for $n > 1$.  Therefore the  
contributions to $H (x, \xi)$ coming from the higher terms $h_{2k} (x)$ in  
(\ref{eq:a3}) die out.

To quantify our observation we perform LO evolution of the off-diagonal 
distributions from a wide variety of different diagonal input distributions.  We present 
a representative selection of our results below.  To be specific we evolve from 
different model $(M)$ inputs based on a set of diagonal distribution at a low scale 
$Q_0$, and show, as the scale $\mu^2 = Q^2$ increases, the speed with which the 
models tend to common off-diagonal distributions
\be
\label{eq:a10}
\hat{H}^M (x, \xi, Q^2) \; \rightarrow \; H (x, \xi, Q^2).
\ee
One possibility is simply to use the diagonal partons themselves as input
\be
\label{eq:a9}
\hat{H}^M (x, \xi, Q_0^2) \; = \; h_0 (x) \; = \; f (x),
\ee
where $f_q (x) = q (x)$ and $f_g (x) = xg (x)$.  We comment on this below.

Of course we can add to the above input distributions, functions with support entirely 
in the time-like ERBL region  $| x | < \xi$.  
Evolution never pushes such functions out 
of the ERBL domain.  On the other hand as $\xi \rightarrow 0$ such contributions 
become invisible and cannot be generated from diagonal parton distributions.  
However here we discuss functions which are analytic in the whole interval $-1 \leq x 
\leq 1$.  So far the most realistic models for the nucleon (such as \cite{POLY}) are 
indeed analytic in the whole interval, and do not contain any $\theta (\xi - | x | )$ 
contributions.

For illustration we show results based on two 
different sets of diagonal partons.  In all cases we start the off-diagonal evolution from 
$Q_0^2 = 0.26$~GeV$^2$ and show results at $Q^2 = 4$ and 100~GeV$^2$.  For set 
1 we take the GRV(98) partons \cite{GRV} and for set 2 the toy model
\be
\label{eq:a11}
xg \; = \; 3 (1 - x)^3, \quad xq \; = \; 9x (1 - x)^2, \quad x \bar{q} \; = \; 0,
\ee
which satisfies the momentum and $n_f = 3$ flavour sum rules.  Figs.~2 and 3 show 
the off-diagonal distributions obtained from GRV-based inputs at $\xi = 0.3$ and $\xi 
= 0.03$ respectively, whereas Figs.~4 and 5 show the corresponding distributions 
based on the toy model partons.  In each plot the {\it dotted curve} corresponds to 
evolution from (unmodified) diagonal partons, as in (\ref{eq:a9}).  We see such an 
approach gives unphysical input forms in the ERBL region $| x | < \xi$, particularly 
for the GRV quark distribution which is singular as $x \rightarrow 0$, that is $f_q (x) 
= q (x) \sim x^{-a}$ with $a > 0$.  It is more realistic to smooth the input in the 
ERBL region.  We show results for two types of smoothing, both of which retain the 
constraints of the polynomial condition.  First the {\it continuous curves} on 
Figs.~2--5 are obtained by modifying the diagonal input by requiring the conformal 
moments (\ref{eq:a5}) of $H (x, \xi, Q_0^2)$ to be independent of $\xi$.  Second 
we modify the diagonal input by using different forms of $h (y)$ in the double 
distribution of ref.~\cite{DD}.  The {\it dashed curves} are the result of one 
interesting choice of $h (y)$.  Before we discuss the results shown in Figs.~2--5 we 
explain, in turn, the two smoothing procedures.

In the first case we seek an input which has the diagonal limit $H (x, \xi = 0; Q_0^2) 
= f (x, Q_0^2)$, but whose conformal moments do not depend on $\xi$, that is $O_n 
(\xi) = O_n (0)$.  This input may be calculated by the Shuvaev transform \cite{S1,S2}
\be
\label{eq:a12}
H (x, \xi) \; = \; \int_{-1}^1 \: dx^\prime \left [ \frac{2}{\pi} \: {\rm Im} \: \int_0^1 \: 
ds \: \frac{(x + \xi (1 - 2s))^p}{y (s) \sqrt{1 - y (s) x^\prime}} \right ] \: 
\frac{d}{dx^\prime} \: \left ( \frac{f (x^\prime)}{x^{\prime p} | x^\prime |} \right )
\ee
with $p = 1$ for gluons and $p = 0$ for quarks, where
$$
y (s) \; = \; \frac{4s (1 - s)}{x + \xi (1 - 2s)}.
$$
For small values of $\xi$ and $x$ this prescription will in fact generate reliable 
off-diagonal distributions from known diagonal partons {\it at any scale} \cite{S2}.  
For large $\xi$ it should produce a more physical input than the unmodified diagonal 
form (\ref{eq:a9}), and will much more quickly evolve into the true off-diagonal 
distributions.  The results are shown by the continuous curves in Figs.~2--5; the 
smoothing of the distributions in the ERBL-region is manifest.

An alternative way to smooth out the input distributions is to use the double 
distributions of ref.~\cite{DD}
\be
\label{eq:a13}
H (x, \xi) \; = \; \int_a^b \: F (x - \xi y, y) \: dy
\ee
with
\be
\label{eq:a14}
F (x, y) \; = \; f (x) \: h (y) \: \bigg{/} \: 2 \int_0^{1 - x} \: h (y^\prime) \: dy^\prime\;,
\ee
where $h (y)$ is an even function of $y$ and $f(x)$ is a diagonal distribution.
Thus in the limit $\xi=0$ we have $H (x, \xi = 0) = f (x)$.
The integration limits in (\ref{eq:a13}) are
\bea
\label{eq:106}
(a,b)\;=\;
\left\{  
\begin{array}{ll} 
\left (- \frac{1-x}{1+\xi}, \frac{1-x}{1-\xi} \right)   
& \mbox{~~~~~for~~~~~$x>\xi$}  \\ \\
\left (- \frac{1-x}{1+\xi}, \frac{1+x}{1+\xi} \right)   
& \mbox{~~~~~for~~~ $-\xi<x<\xi$ }  \\ \\
\left (- \frac{1+x}{1-\xi}, \frac{1+x}{1+\xi} \right)   
& \mbox{~~~~~for~~~~~$x<-\xi$}\;.  \\ 
\end{array} 
\right. 
\eea
This construction satisfies the polynomial condition. Indeed,
if we write relation (\ref{eq:a13}) in the equivalent form
\be
\label{eq:101}
H(x,\xi) \;=\; \int_R d\xtil\, dy \;
F(\xtil, y)\; \delta(x-(\xtil+\xi y))\;
\ee
where the integration region $R$ is given by the square in the $(\xtil,y)$ plane
defined by the vertices: $(0,\pm 1)$ and $(\pm 1,0)$, then condition
(\ref{eq:a2}) is explicitly satisfied. 

We are free to choose the form of $h (y)$ and generate a whole range of input 
distributions.  The choice
\be
\label{eq:a15}
h (y) \; = \; (1 - y^2)^{p + 1},
\ee
with $p = 0, 1$ for quark (gluons), generates an input form which turns out 
remarkably similar to that obtained by the Shuvaev transform (\ref{eq:a12}).  We 
tried several other forms for $h (y)$.  A relevant choice is
\be
\label{eq:a16}
h (y) \; = \; \sin (\pi y^2)
\ee
for both quarks and gluons, since it generates an oscillatory input behaviour in the 
ERBL region which is qualitatively similar to the quark distribution calculated in 
ref.~\cite{POLY}.  However after evolving to $Q^2 = 4$~GeV$^2$ we see that the 
oscillations are much weaker, and that by $Q^2 = 100$~GeV$^2$ they have already 
disappeared.

The similarity of the input based on (\ref{eq:a12}) and (\ref{eq:a15}) at scale $Q_0^2 
= 0.26$~GeV$^2$ is fortuitous.  The Shuvaev transform (\ref{eq:a12}) has the 
advantage that it is based on conformal moments which renormalize multiplicatively, 
(\ref{eq:a8}), with the same anomalous dimensions $\gamma_n$ as in the diagonal 
case.  So the transform is stable to a change of scale, whereas this is not the case for 
(\ref{eq:a15}).  Therefore prescriptions based on (\ref{eq:a12}) and (\ref{eq:a15}) 
will give different off-diagonal distributions if applied at a higher scale.

From the representative results shown in Figs.~2--5, and from those obtained by other 
trial inputs based on diagonal distributions, we can draw the following conclusions.
\begin{itemize}
\item[(i)] If we are given a set of diagonal partons $f (x, Q_0^2)$ then we can predict 
the off-diagonal distributions $H (x, \xi, Q^2)$ at a higher scale with a high degree of 
certainty.  Any differences, between the prescriptions used, rapidly disappear with 
increasing $Q^2$.
\item[(ii)] The prescriptions are most precise at small values of $\xi$, as expected 
from ref.~\cite{S2}.
\item[(iii)] There is freedom in the form of the input distributions in the ERBL region, 
$| x | < \xi$, but even here the various choices evolve to a common limit reasonably 
rapidly with increasing $Q^2$, as shown, for example, by comparing the continuous 
and dashed curves in Figs.~2--5.  (The dotted curves, corresponding to unmodified 
input partons, are shown simply for comparison.)
\item[(iv)] For gluons in the whole domain, and for quarks in 
the DGLAP region $| x | > \xi$, 
the distributions evolve particularly quickly to a common limit.
\end{itemize}

These conclusions are encouraging for phenomenology.  For instance they will enable 
data for processes described by off-diagonal distributions to provide additional 
independent constraints in the global (diagonal) parton analyses.  For example, one 
type of process for which data are becoming available is high energy diffractive vector 
meson electroproduction.  The description is dependent on the off-diagonal gluon 
distribution at small $x, \xi$, which is especially well known in terms of the diagonal 
gluon distribution, see Figs.~3 and 5. \\

\noindent {\large \bf Acknowledgements}

This work was supported in part by the Royal Society, INTAS (95-311), the Russian 
Fund for Fundamental Research (98~02~17629), the Polish State Committee for 
Scientific Research grant No. 2 P03B~089~13 and by the EU Fourth Framework 
Programme TMR, Network \lq QCD and the Deep Structure of Elementary 
Particles\rq~contract FMRX-CT98-0194 (DG12-MIKT) and (DG12-MIHT).

\newpage  
\begin{figure}  
   \vspace*{-1cm}  
    \centerline{  
     \epsfig{figure=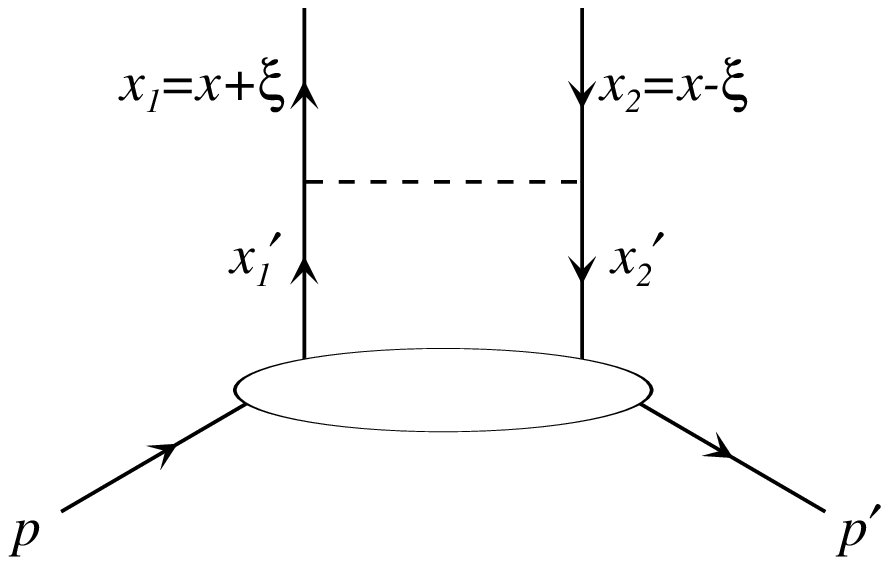,width=15cm}  
               }  
    \vspace*{-0.5cm}  
\caption{A schematic diagram showing the variables for the off-diagonal parton 
distribution $H (x, \xi)$, where $x_{1,2} = x \pm \xi$.  The timelike region with $| x | 
< \xi$ is frequently called the ERBL domain, since in the limit $\xi \rightarrow 1$ 
\lq\lq pure\rq\rq~ERBL evolution applies.  Similarly $| x | > \xi$ is known as the 
DGLAP domain since we have \lq\lq pure\rq\rq~DGLAP evolution for $\xi 
\rightarrow 0$. 
}  
\label{fig1}  
\end{figure} 

\newpage  
\begin{figure} 
   \vspace*{-1cm} 
    \centerline{ 
     \epsfig{figure=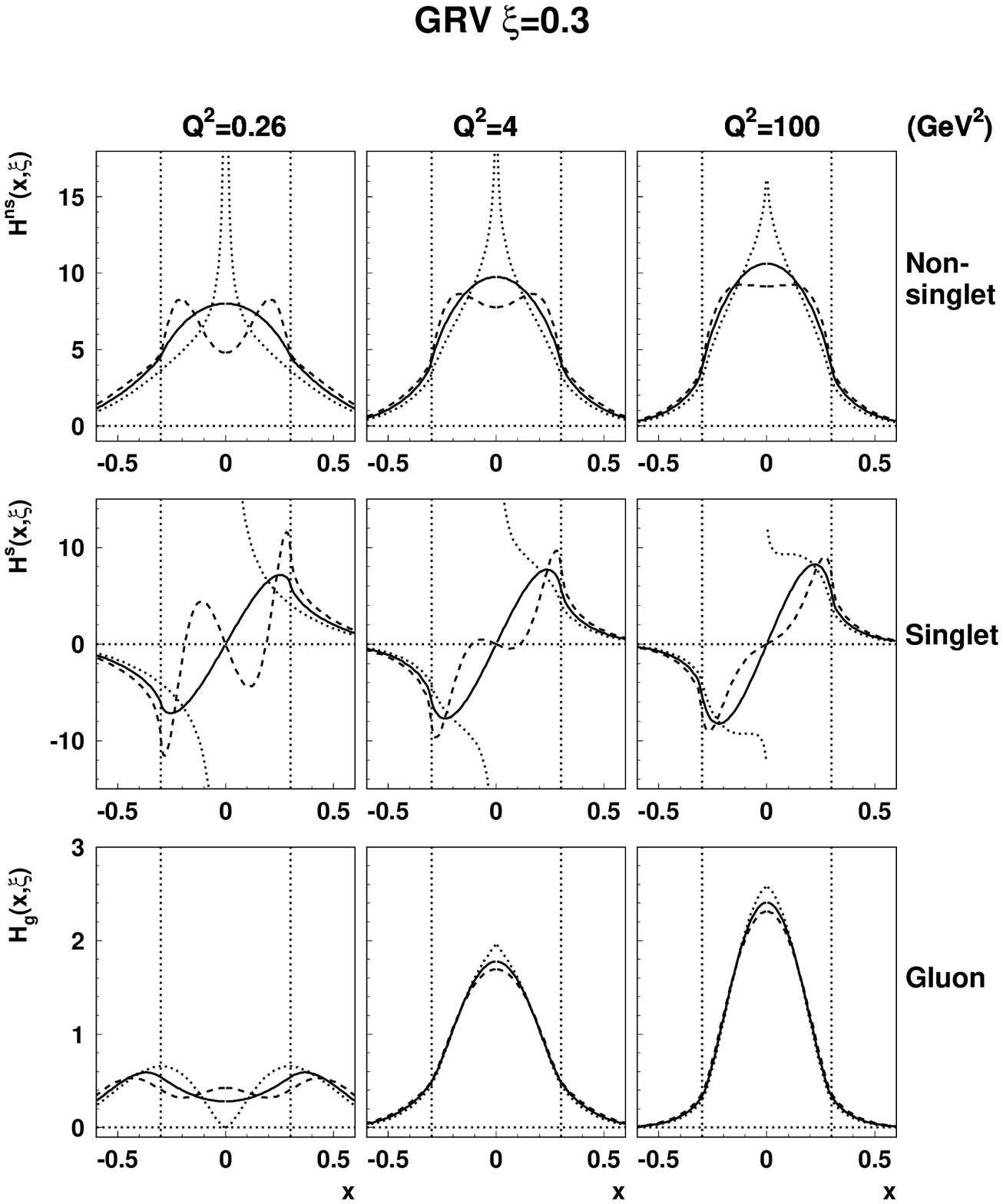,width=18cm}
             } 
\vspace*{-0.5cm}  
\caption{The evolution to $Q^2 = 4$ and 100~GeV$^2$ of the quark non-singlet 
$H^{ns}$, the quark singlet $H^s$ and the gluon distribution $H_g (= x H_g^{\rm 
Ji})$ starting from diagonal GRV-based input at $Q_0^2 = 0.26$~GeV$^2$ for $\xi = 
0.3$.  The dotted curves correspond to evolution from unmodified input partons as in 
(\ref{eq:a9}), the continuous curves to using the \lq\lq Shuvaev prescription\rq\rq~of 
(\ref{eq:a12}) and the dashed curves to \lq\lq double distribution\rq\rq~input with 
(\ref{eq:a16}).         
} 
\label{fig2} 
\end{figure}  

\newpage  
\begin{figure} 
   \vspace*{-1cm} 
    \centerline{ 
     \epsfig{figure=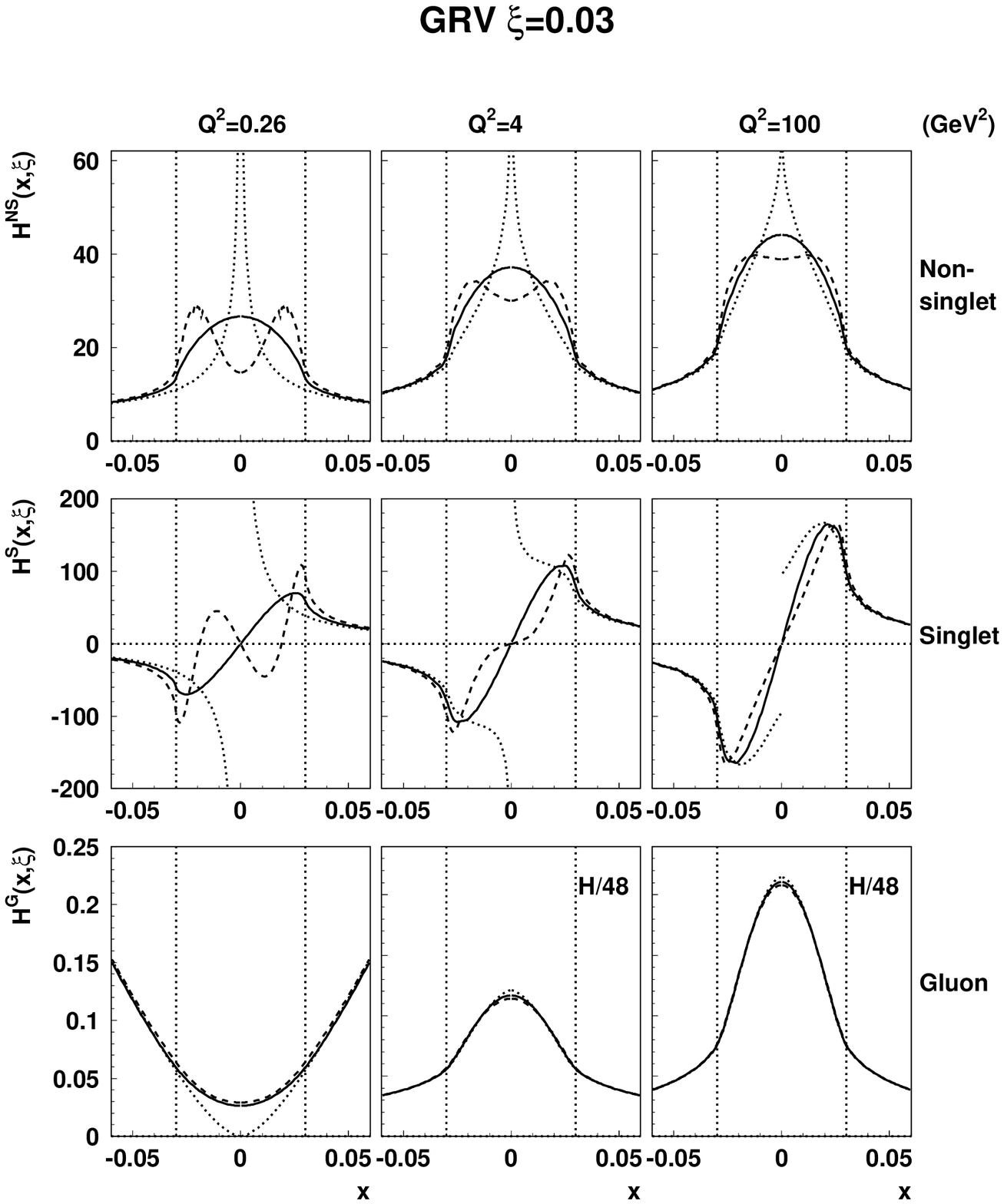,width=18cm} 
               } 
    \vspace*{-0.5cm} 
\caption{As in Fig.~2 but for $\xi = 0.03$.} 
\label{fig3} 
\end{figure}

\newpage   
\begin{figure}  
   \vspace*{-1cm}  
    \centerline{  
     \epsfig{figure=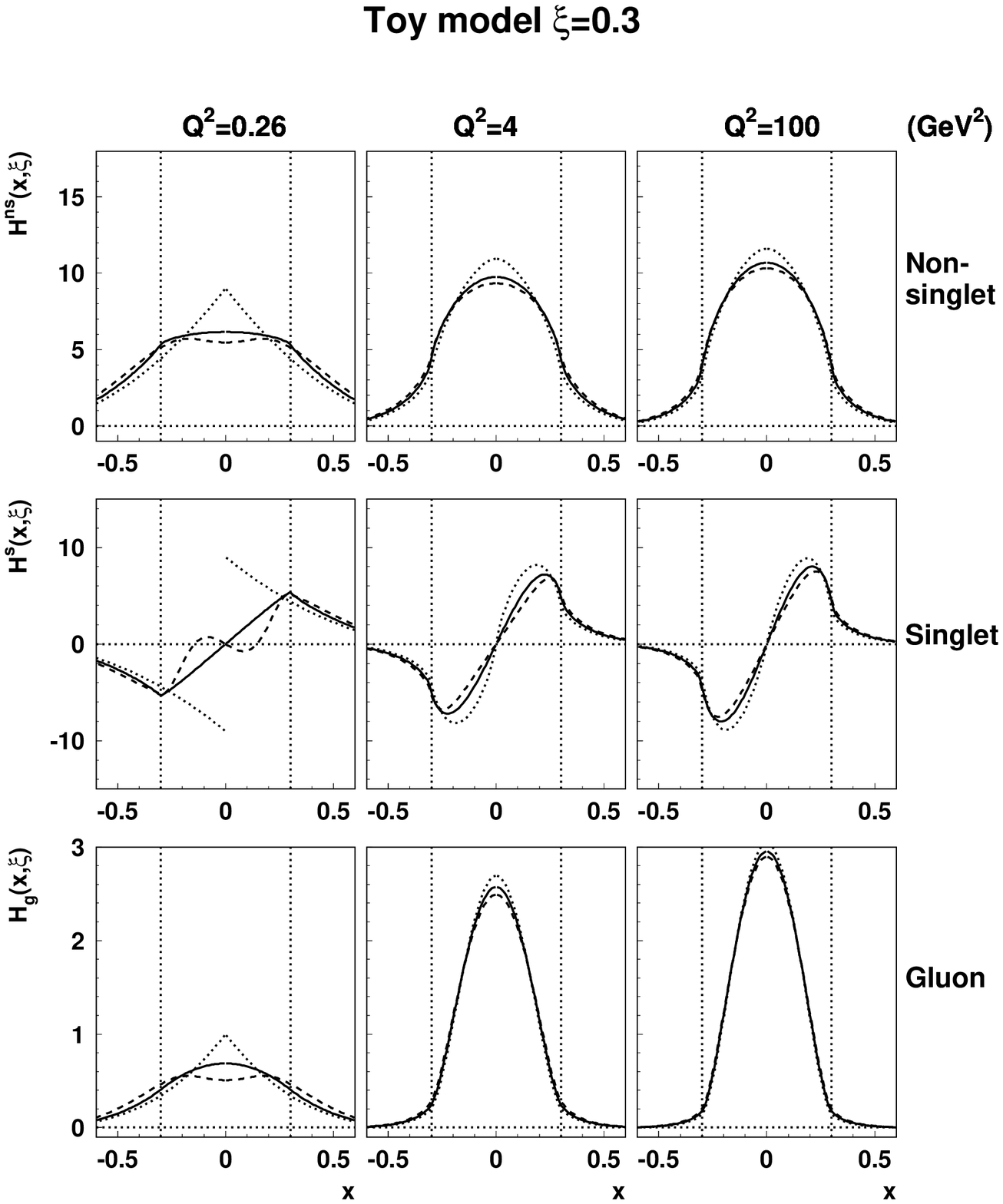,width=18cm}  
               }  
    \vspace*{-0.5cm}  
\caption{As in Fig.~2 but based on the toy model diagonal partons of 
(\ref{eq:a11}); $\xi = 0.3$.}  
\label{fig4}  
\end{figure}

\newpage   
\begin{figure}  
   \vspace*{-1cm}  
    \centerline{  
     \epsfig{figure=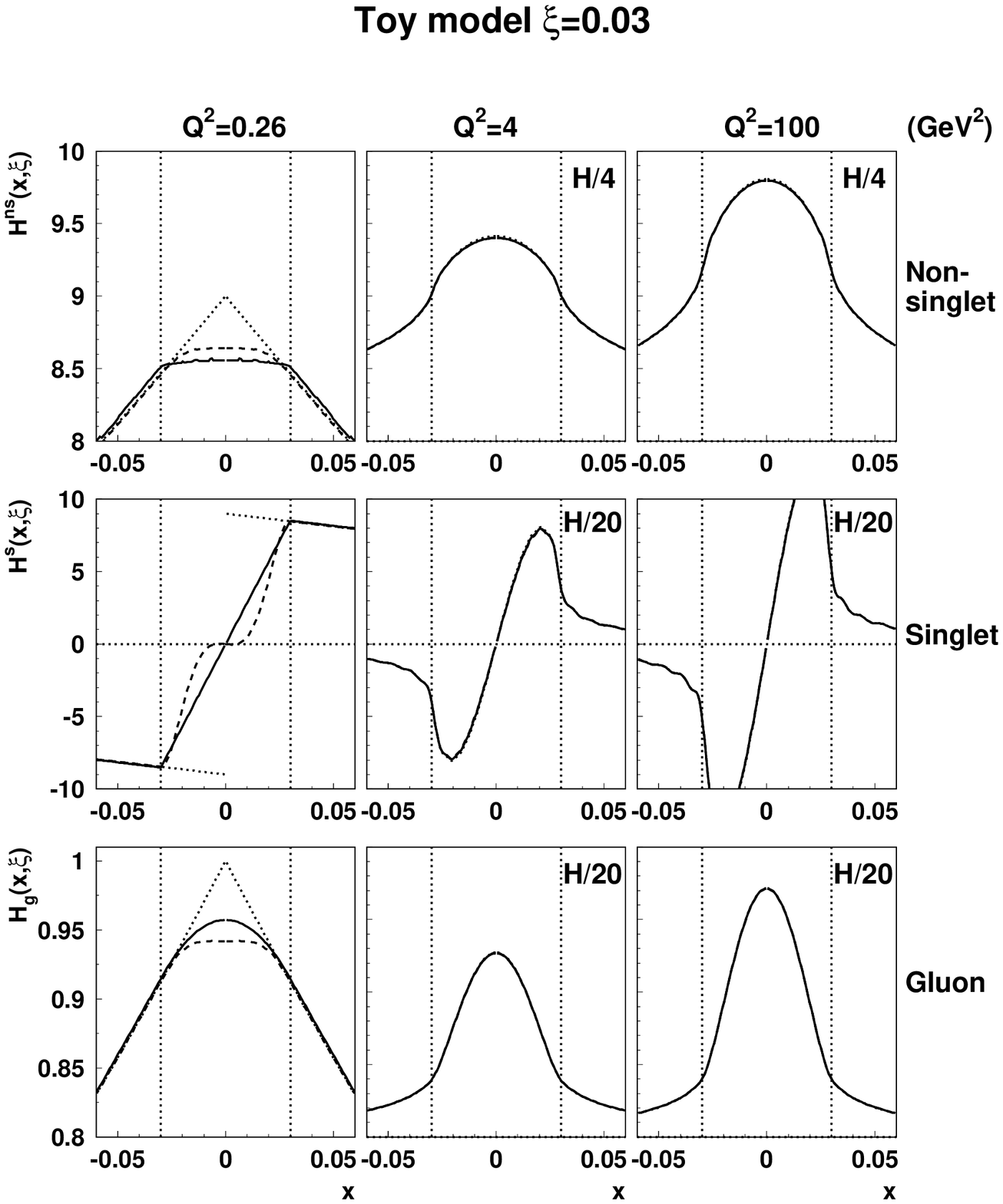,width=18cm}  
               }  
    \vspace*{-0.5cm}  
\caption{As in Fig.~4 but for $\xi = 0.03$.}  
\label{fig5}  
\end{figure}

\end{document}